\RequirePackage{latexrelease}
\documentclass[twocolumn,english,superscriptaddress,citeautoscript,preprintnumbers,amsmath,amssymb,prl,floatfix,footinbib]{revtex4-2}
\usepackage{hyperref}
\hypersetup{colorlinks=true,linkcolor=red,citecolor=blue}
\usepackage[scaled=2]{helvet}
\usepackage[T1]{fontenc}
\usepackage[latin9]{inputenc}
\usepackage{array}
\usepackage{float}
\usepackage{multirow}
\usepackage{amsmath}
\usepackage{amssymb}
\usepackage{graphicx}
\usepackage{csquotes}
\usepackage{xcolor,soul}
\usepackage{dsfont}

\makeatletter

\providecommand{\tabularnewline}{\\}

\@ifundefined{textcolor}{}
{%
	\definecolor{BLACK}{gray}{0}
	\definecolor{WHITE}{gray}{1}
	\definecolor{RED}{rgb}{1,0,0}
	\definecolor{GREEN}{rgb}{0,1,0}
	\definecolor{BLUE}{rgb}{0,0,1}
	\definecolor{CYAN}{cmyk}{1,0,0,0}
	\definecolor{MAGENTA}{cmyk}{0,1,0,0}
	\definecolor{YELLOW}{cmyk}{0,0,1,0}
}

\usepackage[scaled=2]{helvet}\usepackage{amstext}

\makeatletter

\usepackage{dcolumn}
\usepackage{bm}
\usepackage{times}

\makeatother

\usepackage{babel}

\makeatother
\usepackage{babel}
\begin{document}
	\title{Anomalous Hall effect and magnetic structure of the topological semimetal - (Mn$_{0.78}$Fe$_{0.22}$)$_{3}$Ge}
	\author{V. Rai}
	\affiliation{Forschungszentrum J\"ulich GmbH, J\"ulich Centre for Neutron Science (JCNS-2)
		and Peter Gr\"unberg Institut (PGI-4), JARA-FIT, 52425 J\"ulich, Germany}
	\affiliation{RWTH Aachen, Lehrstuhl f\"ur Experimentalphysik IVc, J\"ulich-Aachen Research
		Alliance (JARA-FIT), 52074 Aachen, Germany}
	\author{A. Stunault}
	\affiliation{Institut Laue-Langevin, 71 Avenue des Martyrs, 38042 Grenoble Cedex 9, France}
	\author{W. Schmidt}
	\affiliation{Forschungszentrum J\"ulich GmbH, J\"ulich Centre for Neutron Science at ILL,
		71 Avenue des Martyrs, 38042 Grenoble, France}
	\author{S. Jana}
	\affiliation{Forschungszentrum J\"ulich GmbH, J\"ulich Centre for Neutron Science (JCNS-2)
		and Peter Gr\"unberg Institut (PGI-4), JARA-FIT, 52425 J\"ulich, Germany}
	\affiliation{RWTH Aachen, Lehrstuhl f\"ur Experimentalphysik IVc, J\"ulich-Aachen Research
		Alliance (JARA-FIT), 52074 Aachen, Germany}
	\author{J. Per\ss{}on}
	\affiliation{Forschungszentrum J\"ulich GmbH, J\"ulich Centre for Neutron Science (JCNS-2)
		and Peter Gr\"unberg Institut (PGI-4), JARA-FIT, 52425 J\"ulich, Germany}
	\author{J.-R. Soh}
	\affiliation{Institute of Physics, Ecole Polytechnique F\'ed\'erale de Lausanne (EPFL), CH-1015 Lausanne, Switzerland}
	\author{Th. Br\"uckel}
	\affiliation{Forschungszentrum J\"ulich GmbH, J\"ulich Centre for Neutron Science (JCNS-2)
		and Peter Gr\"unberg Institut (PGI-4), JARA-FIT, 52425 J\"ulich, Germany}
	\affiliation{RWTH Aachen, Lehrstuhl f\"ur Experimentalphysik IVc, J\"ulich-Aachen Research
		Alliance (JARA-FIT), 52074 Aachen, Germany}
	\author{S. Nandi}
	\email{s.nandi@fz-juelich.de}
	\affiliation{Forschungszentrum J\"ulich GmbH, J\"ulich Centre for Neutron Science (JCNS-2)
		and Peter Gr\"unberg Institut (PGI-4), JARA-FIT, 52425 J\"ulich, Germany}
	\affiliation{RWTH Aachen, Lehrstuhl f\"ur Experimentalphysik IVc, J\"ulich-Aachen Research
		Alliance (JARA-FIT), 52074 Aachen, Germany}
	\begin{abstract}
		Me$_{3+\delta}$Ge, being a Weyl semimetal, shows a large anomalous Hall effect (AHE), which decreases slowly with an increase in $\delta$ from 0.1 to 0.4. However, AHE in this compound remains significantly large in the whole range of $\delta$ because of the robust nature of the topology of bands. To explore the possibility of tuning the anomalous transport effects in Weyl semimetals, we have studied the single-crystal hexagonal-(Mn$_{0.78}$Fe$_{0.22}$)$_3$Ge compound. Magnetization of this compound shows two magnetic transitions at 242 K ($T_{\text{N1}}$) and 120 K ($T_{\text{N2}}$). We observed that the AHE persists between $T_{\text{N2}}$ - $T_{\text{N1}}$ and vanishes below $T_{\text{N2}}$. Further, we performed single-crystal neutron diffraction experiments (using spherical neutron polarimetry and unpolarized neutron diffraction) to determine the magnetic structures of (Mn$_{0.78}$Fe$_{0.22}$)$_3$Ge at different temperatures. Our neutron diffraction results show that the sample possesses a collinear antiferromagnetic structure below $T_{\text{N2}}$. However, the magnetic structure of the sample remains noncollinear antiferromagnetic, the same as Mn$_3$Ge, between $T_{\text{N1}}$ to $T_{\text{N2}}$. The presence of AHE, and noncollinear magnetic structure in (Mn$_{0.78}$Fe$_{0.22}$)$_3$Ge, between $T_{\text{N1}}$ and $T_{\text{N2}}$, suggest the existence of Weyl points in this temperature regime. Below $T_{\text{N2}}$,  AHE is absent, and the magnetic structure also changes to a collinear antiferromagnetic structure. These observations signify a strong link between the magnetic structure of the sample and AHE. 
	\end{abstract}
	\maketitle

	\section{Introduction:}
	Chiral antiferromagnetic Weyl semimetals have gained enormous attention due to the strong Berry curvature, which generates a large intrinsic magnetic field (in phase space), leading to the observation of anomalous Hall effect, Nernst effect, and other transport effects even in the absence of the external magnetic field \hbox{\cite{yang2017topological, kiyohara2016giant, kubler2014non, chen2014anomalous, kuroda2017evidence, chen2021anomalous}}. A combination of such effects is known as anomalous transport effects. Among these anomalous transport effects, the anomalous Hall effect (AHE) gains special attention as it is easier to measure and provides evidence for the intrinsic magnetic field present in the system {\cite{kiyohara2016giant}}. Compared to ferromagnetic (FM) materials, which also have large AHE \cite{nagaosa2010anomalous, sales2006anomalous, meng2019large}, chiral antiferromagnetic (AFM) materials do not suffer from large stray magnetic fields and large Hall hysteresis. As a result, they are more attractive for spintronic applications due to the potential for downscaling and increased efficiencies.
	
	Hexagonal-Mn$_3$Ge is a well-known magnetic Weyl semimetal driven by time-reversal symmetry (TRS) breaking \mbox{\cite{soh2020ground, yang2017topological,kubler2014non, wu2020magneto, chen2021anomalous, wuttke2019berry, nayak2016large}}. The noncollinear AFM structure of Mn$_3$Ge leads to non-zero Berry curvature, resulting in an intrinsic magnetic field equivalent to about 200 T {\cite{kiyohara2016giant}}, leading to the observation of a large AHE at low temperatures.  The AHE in this compound can be observed in the temperature range of 2 K - 365 K \cite{nayak2016large, kiyohara2016giant,rai_mn3ge}. The recent discovery of the conversion of singlet spin-unpolarized Cooper pairs into triplet spin-polarized pairs in the epitaxial thin film of Mn$_3$Ge provides unique experimental evidence of the presence of the intrinsic magnetic field in Mn$_3$Ge \cite{jeon2021long}. The Hall hysteresis in Mn$_3$Ge remains  small ($<200$ Oe), and its Hall resistivity can be reversed easily using a small magnetic field ($<100$ Oe) \cite{nayak2016large, kiyohara2016giant, rai_mn3ge}. These electrical transport features have led to enormous attention in the field of thin-film and single-crystal Mn$_3$Ge, and similar Weyl-semimetals \cite{jeong2016termination, dung2011magnetism, hong2020large, wang2021robust, jeon2021long}, which can accelerate the development of efficient magnetic switching, memory or other spintronic devices.
	
	To utilize the Weyl semimetals for application purposes, it is important to determine the parameters which can tune Weyl points and resulting transport effects. We have explored the Fe doping in the hexagonal-Mn$_3$Ge compound to study the effect of doping on AHE. Fe doping in Mn$_3$Ge has been studied a long time back \cite{lecocq1963, kanematsu1967,  hori1992antiferromagnetic,hori1995magnetic, niida1995magnetic,  du2007abnormal, du2007giant, albertini2004magnetocrystalline}. It has been reported in Refs. \cite{hori1992antiferromagnetic, niida1995magnetic} that the N\'eel temperature ($T_{\text{N}}$) of hexagonal-Mn$_3$Ge decreases with Fe doping and vanishes near $30\%$ Fe doping. An additional magnetic transition below $T_{\text{N}}$ also appears for compounds between 15 - 30\% Fe doping. Above $\sim$ 30$\%$ Fe doping, both the transitions are suppressed, and the sample becomes ferromagnetic \mbox{\cite{hori1992antiferromagnetic, niida1995magnetic}}. Mn$_3$Ge (polycrystalline) compounds with nearly 17-23\% Fe doping have gained special attention because of multiple magnetic transitions, abnormal magnetoresistance, and large magnetocaloric effect \cite{du2007abnormal,du2007giant}. However, the topological aspects of these materials have not been studied yet. Therefore, we explored the magnetic and transport properties of the 22\% Fe doped Mn$_3$Ge to determine the change in magnetic properties and AHE with Fe doping in Mn$_3$Ge.
	
	In this report, we first discuss the magnetization of the single-crystal (Mn$_{0.78}$Fe$_{0.22}$)$_3$Ge below 300 K, where magnetic phase transitions were observed at 242 K ($T_{\text{N1}}$) and 120 K ($T_{\text{N2}}$). Following this, we report the Hall effect measurements below 300 K, where a non-zero anomalous Hall effect has been observed between $T_{\text{N1}}$ and $T_{\text{N2}}$. Further, we discuss the neutron diffraction measurements to determine the magnetic structures of (Mn$_{0.78}$Fe$_{0.22}$)$_3$Ge below $T_{\text{N2}}$, and between $T_{\text{N2}}$ and $T_{\text{N1}}$. We have determined the magnetic structures of the compound at different temperatures using unpolarized and polarized neutron diffraction experiments. Our neutron diffraction analyses show that, at 4 K, the sample is AFM with Mn+Fe moments lying along the \textit{c} axis. In this magnetic phase, Mn+Fe moments are ferromagnetically arranged within the \textit{a-b} plane. However, the direction of Mn+Fe moments located at $c=0.25$ and $c=0.75$ are opposite. In contrast to this, at 130 K, the sample possesses the same magnetic structure as that of Mn$_3$Ge, which has a noncollinear AFM structure in the \textit{a-b} plane {\cite{soh2020ground}}.  The magnetic structures in different temperature regimes are also consistent with our magnetization results.  It is an interesting finding that the AHE in our compound is observed only in the temperature regime where its magnetic structure remains the same as the parent phase Mn$_3$Ge, which is an established Weyl semimetal. This suggests that the (Mn$_{0.78}$Fe$_{0.22}$)$_3$Ge is also a Weyl semimetal, which indicates the robust nature of the topology of the electronic bands.  However, further experimental studies such as ARPES (angle-resolved photoemission spectroscopy) and theoretical evidence are required to confirm the presence and location of the Weyl points with respect to the Fermi surface.  The persistence of Weyl points in a significantly doped sample could be of great importance, as it opens an opportunity for tuning the various  anomalous transport effects.

	\section{Experimental details} \label{exp_detail}
	
	The single crystals used for the experiment were prepared using the self-flux method.  The phase diagram (Ref. {\cite{binary1}} )  indicates that the parent sample (Mn$_3$Ge) stabilizes in the presence of an excess of Mn. Therefore, we have synthesized single-crystals of hexagonal-(Mn$_{0.78}$Fe$_{0.22}$)$_{3.2}$Ge for our study. To prepare the sample, high purity Mn, Fe, and Ge were taken in the stoichiometric ratio of Mn:Fe:Ge = 2.5:0.8:1. The elements were melted using the induction melting technique, and sealed into a quartz tube. Further, it was heated into the furnace at 1293 K for 5 hrs, followed by slow cooling (2 K/hr) down to 1123 K. The hexagonal phase of Mn$_3$Ge remains stable above 920 K and metastable below this temperature. Therefore, the quartz tube containing the sample was quenched into the water at 1123 K to retain the high-temperature hexagonal phase. Chemical analysis of the sample was performed using the inductively coupled plasma optical emission spectroscopy (ICP-OES) method. Assuming that the Fe replaces the Mn atoms, the empirical formula of the sample was found to be (Mn$_{0.79(1)}$Fe$_{0.21(1)}$)$_{3.21(8)}$Ge. For convenience, we will refer to the compound as (Mn$_{0.78}$Fe$_{0.22}$)$_{3}$Ge. Single-crystal X-ray diffraction (SC-XRD) of a small crystal piece ($\sim$ 100 $\mu$m) was performed using the Rigaku Oxford Diffraction (SuperNova) instrument (not shown), which displayed sharp Bragg reflections with a hexagonal reciprocal space pattern, confirming the high quality of the sample. X-ray powder diffraction (XRPD) of the sample was performed (not shown) using the Huber Imaging Plate Guinier Camera G670 instrument. The data were analyzed using the FullProf software \cite{rodriguez1993recent}, and a fitting parameter, $\chi^2=1.77$ was observed. The analysis showed that the sample synthesized in a single (hexagonal) phase with the $P6_3/mmc$ space group symmetry, the same as the parent compound hexagonal-Mn$_3$Ge. The lattice parameters at room temperature were found to be \textit{a} = \textit{b} = 5.290(5) {\AA}, \textit{c} =  4.297(4) {\AA}, which are slightly smaller than the lattice parameters Mn$_3$Ge \cite{kiyohara2016giant} and similar to the values reported by Refs. \cite{niida1995magnetic, kanematsu1967}.

	\begin{figure}
		\includegraphics[width=8cm]{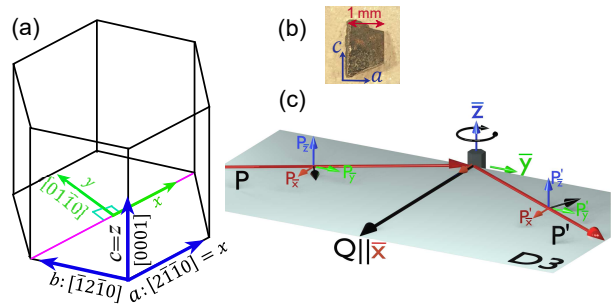} \caption{(a) Crystallographic axes \textit{a, b}, and \textit{c} are shown in the hexagonal coordinate system. On the other hand, \textit{x, y}, and  \textit{z} axes, which are shown relative to the \textit{a, b}, and \textit{c}, lie in the Cartesian coordinate system. (b) Picture of the single-crystal used for the neutron diffraction experiments. The axes \textit{a} and \textit{c} were determined using the Laue diffraction. (c) The geometry of the D3+CRYOPAD experimental setup in the horizontal (scattering) plane. The sample is rotated about the vertical direction ($\bar{\mathsf{z}}$ axis). $\mathbf{P}$ and $\mathbf{P}'$ denote incoming and scattered neutron beam polarization, respectively. $P_i$ ($i=\bar{\mathsf{x}}, \bar{\mathsf{y}}, \bar{\mathsf{z}}$), and $P_j'$ ($j=\bar{\mathsf{x}}, \bar{\mathsf{y}}, \bar{\mathsf{z}}$) denote different polarization components of the incoming and scattered beam. $\bar{\mathsf{x}}, \bar{\mathsf{y}}$, and $\bar{\mathsf{z}}$  form a Cartesian coordinate system.}
		\label{fig:exp_all}
	\end{figure}

	\begin{figure*}
		\includegraphics[width=17cm]{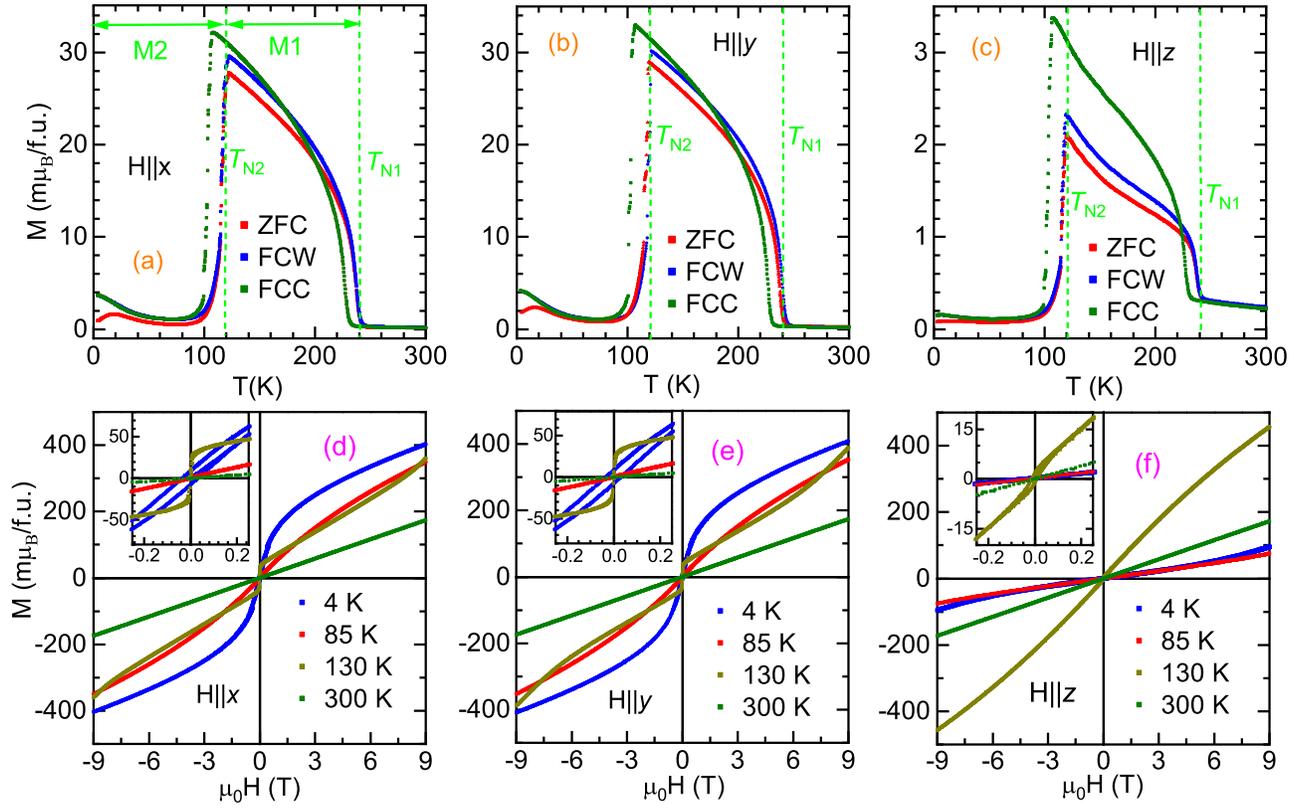} \caption{(a-c) The temperature-dependent magnetization (\textit{M}) of the doped sample with 100 Oe magnetic field ({$\mu_0H$}) applied along the \textit{x, y,} and \textit{z} axes, respectively. ZFC signifies zero-field cooling condition. FCC and FCW mean the data were collected while cooling and warming the sample, respectively. (d-f) The magnetic field-dependent magnetization (\textit{M}(\textit{H})) of the sample, at various temperatures, along the \textit{x, y} and \textit{z} axes. The magnified version of the \textit{M}(\textit{H}) along the \textit{x, y}, and  \textit{z} axes are shown in the inset of (d), (e), and (f), respectively.}
		\label{fig:mt_mh_all}
	\end{figure*}

	All the measurements were performed using the single-crystals oriented with the help of the Laue Camera. Fig. \ref{fig:exp_all}(a) shows crystallographic axes in Cartesian and hexagonal coordinate systems. The axes \textit{x} and \textit{y} lie along the [$2\bar{1}\bar{1}$0], [$01\bar{1}0$] directions, respectively. \textit{z} axis lies parallel to the \textit{c}, which is along the [0001] direction of the hexagonal lattice. The magnetization and electrical transport measurements of the sample were performed using Quantum Design Physical Property Measurement (QD-PPMS) system and Quantum Design DynaCool (QD-DC) instruments. The details of these measurements will be discussed in the next section.

	The unpolarized neutron diffraction measurement of single-crystal (Mn$_{0.78}$Fe$_{0.22}$)$_3$Ge was performed using the D23 lifting counter diffractometer at the Institute of Laue Langevin (ILL), France {\cite{D23}}.  For this experiment, a single-crystal with $\sim$ 1.3$\times$1 $\times$0.8 mm$^3$ dimension was cut (Fig. \ref{fig:exp_all}(b)) after aligning the axes using the Laue camera. A neutron beam with a monochromatic wavelength of 1.27 {\AA} was selected by the Cu (200) monochromator. The crystal was mounted inside a standard orange cryostat and oriented in the 2-circle geometry. For each of the accessible crystallographic planes, the integrated intensities of several Bragg reflections were measured. All the integrated intensities were corrected by the Lorentz factor in the subsequent data analysis. Other correction factors, such as absorption correction, resolution factor, and secondary extinction factor were found to be small and were ignored during the data analysis. The details of various correction factors can be found in Refs. \mbox{\cite{scattering2012neutron, nandi2014magnetic}}.	
		Initially, integrated intensities of a large set of Bragg reflections were collected at 300 K to determine the crystal structure since the sample is paramagnetic at this temperature. Further, the integrated intensities of the same set of Bragg reflections were recorded at  130 K, 85 K, and 4 K. Additional intensity on the nuclear Bragg positions was observed at these temperatures, which was analyzed to determine the magnetic structure of the sample in different temperature regimes.
	
	The polarized neutron diffraction experiment was performed using the CRYOPAD neutron polarimeter \cite{tasset1988determination,tasset1989zero} installed at the D3 instrument at Institut Laue Langevin (ILL), France. A polarized monochromatic neutron beam with 0.832 {\AA} wavelength was selected using the 
	Cu$_2$MnAl monochromator. The polarization of the scattered neutrons was determined using the $^3$He spin filter. The polarization of the incident neutrons remains at 0.935 throughout the experiment. The same single-crystal was used for both the unpolarized (D23) and polarized (CRYOPAD) neutron diffraction experiments. A small ferromagnetic (FM) moment ($\sim$ 0.028(3) $\mu_{\text{B}}$/f.u.) is present in the sample, which can slightly depolarize the incident neutrons. Therefore, to minimize the depolarization, we used a small crystal weighing $\sim$ 13 mg and having a path length of 0.8 - 1.3 mm (Fig. \ref{fig:exp_all}(b)). 
	
	The CRYOPAD setup used for the experiment is illustrated in Fig. \ref{fig:exp_all}(c). During the experiment, the sample is rotated about the vertical axis which is considered as the $\bar{\mathsf{z}}$ axis. The $\bar{\mathsf{x}}$ axis is defined to be parallel to the scattering vector ($\mathbf{Q}$), and $\bar{\mathsf{y}}$ axis lies perpendicular to the $\bar{\mathsf{x}}$ and $\bar{\mathsf{z}}$ axes, such that $\bar{\mathsf{x}}$, $\bar{\mathsf{y}}$, and $\bar{\mathsf{z}}$ axes follow the conventional right-handed Cartesian system. The polarization of the incoming neutron beam is set along one of the  $\bar{\mathsf{x}}$, $\bar{\mathsf{y}}$, or $\bar{\mathsf{z}}$ axes (denoted as $P_i,$ ($i=\bar{\mathsf{x}}, \bar{\mathsf{y}}, \bar{\mathsf{z}}$)), and all the three polarization components of the scattered beam ($P_j',$ ($j=\bar{\mathsf{x}}, \bar{\mathsf{y}}, \bar{\mathsf{z}}$)) are measured. Using this method, various combinations of $P_i$, and $P_j'$ give 3$\times$3 matrix elements of the polarization matrix - $P_{ij}$. To collect $P_{ij}$ corresponding to the different sets of reflections, the experiment was performed in two parts. First, the crystal was oriented with the \textit{b} axis in the vertical position ($\bar{\mathsf{z}}$ axis), to access the ($h0l$) reflections.  Subsequently, the sample was remounted with the \textit{c} axis vertical to measure the polarization matrix for the ($hk0$) reflections. 
	Before performing the experiments in the ($h0l$) and ($hk0$) plane, the sample was cooled under 1 T of the magnetic field applied along the \textit{b} axis of the sample. Vertical and horizontal magnetic field setups were used to apply the magnetic field (along the \textit{b} axis of the sample) during the experiments performed in ($h0l$) and ($hk0$) planes, respectively.
	As we will discuss in the next section, the sample is slightly FM in the \textit{a-b} plane near 130 K. Therefore, the field cooling of the sample populates the FM domains along the direction of the applied field. This also maximizes a particular AFM domain in the sample. The unequal population of ($120^{\circ}$) AFM domains is necessary to determine the magnetic structure of the sample at 130 K {\cite{brown1990determination}} (detail in section {\ref{cryopad}}).

	\section{Magnetization} \label{magnetization}
	The magnetization (\textit{M}) of the sample was measured in the temperature range of 2 K - 300 K. The data show two magnetic phase transitions while heating the sample, near 242(2) K and 120(1) K, which are referred as $T_\text{N1}$ and $T_\text{N2}$, respectively (Fig. \ref{fig:mt_mh_all}(a)). Similar magnetization has been observed by Refs. \cite{kanematsu1967,  hori1992antiferromagnetic, niida1995magnetic} for $\sim$ 20\% Fe doped Mn$_3$Ge samples as well. As shown in Figs. \ref{fig:mt_mh_all}(a-c), the first magnetic transition is observed near 242 K ($T_\text{N1}$), below which the magnetization increases steadily down to  120 K ($T_{\text{N2}}$). We will refer to this temperature range ($T_{\text{N2}}<T<T_{\text{N1}}$) as the \textbf{M1} regime (Fig. \ref{fig:mt_mh_all}(a)). The second magnetic transition appears near 120 K ($T_\text{N2}$), below which the magnetization reduces drastically along all the axes and then slowly increases with a decrease in temperature down to 2 K. We will refer to this temperature range (2 K $<T<T_{\text{N2}}$) as the \textbf{M2} regime (Fig. \ref{fig:mt_mh_all}(a)).

	\begin{figure*}
		\includegraphics[width=17.5cm]{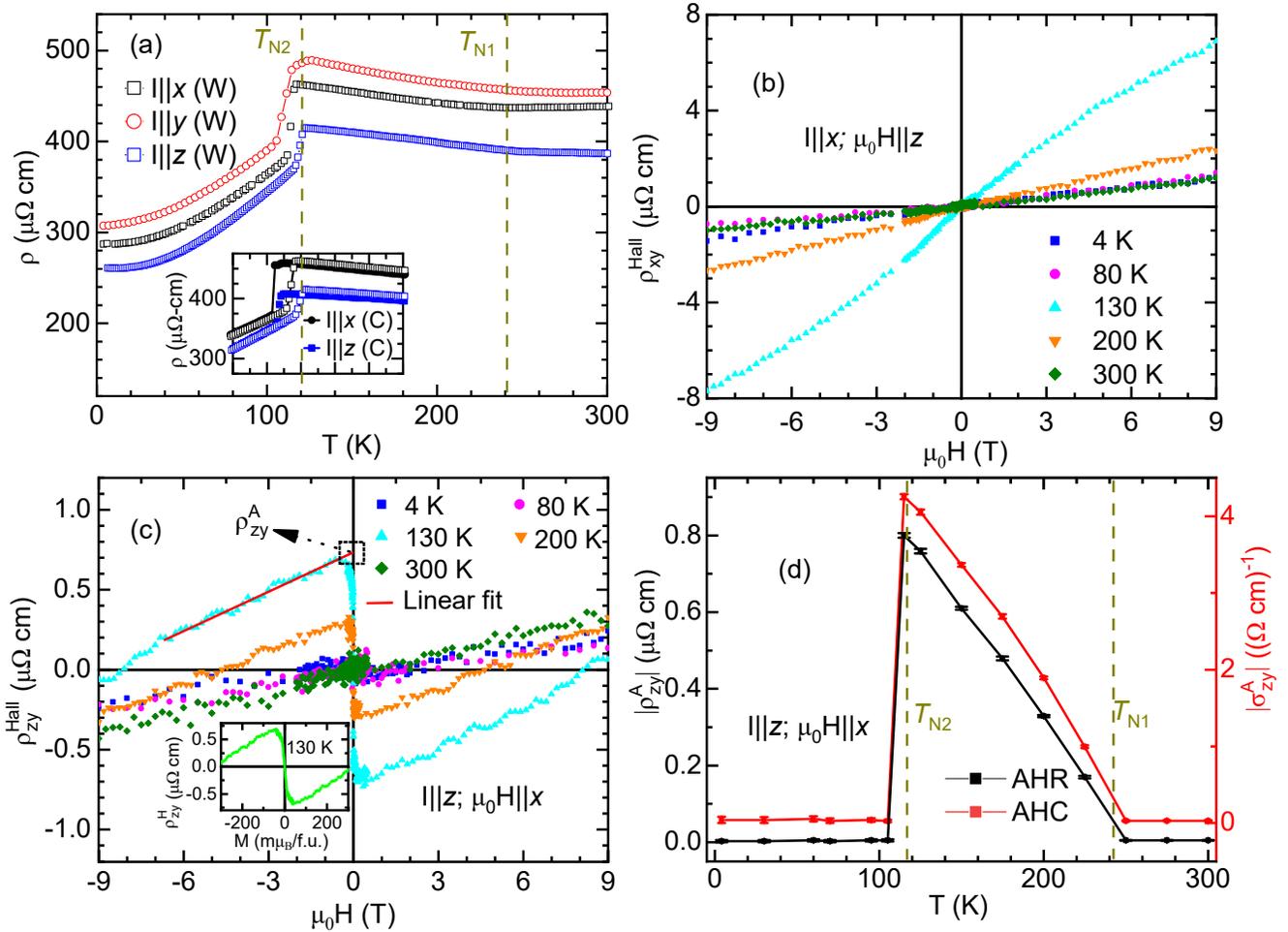}
		\caption{(a) Resistivity of (Mn$_{0.78}$Fe$_{0.22}$)$_3$Ge along the \textit{x, y}, and \textit{z} axes. Inset: Magnified resistivity along the \textit{x} and \textit{z} axes, near 120 K, showing thermal hysteresis. The symbols (C) and (W) correspond to the cooling and warming data, respectively.  (b) Hall resistivity of the sample when the magnetic field is applied along the \textit{z} axis and current is applied along the \textit{x} axis. (c) Hall resistivity ($\rho_{zy}^{\text{Hall}}$) with the magnetic field applied along the \textit{x} axis. $\rho_{zy}^A$ denotes the anomalous  Hall resistivity of the sample, determined by the linear fitting of the high field Hall resistivity data. It is also referred to as anomalous Hall resistivity (AHR) in the main text. Inset: Evolution of the $\rho_{zy}^{\text{Hall}}$ with the magnetization of the sample at 130 K. (d) Temperature dependence of the magnitude of anomalous Hall resistivity (AHR), and anomalous Hall conductivity (AHC) is shown. AHR and AHC are also denoted as $\rho_{zy}^A$ and $\sigma_{zy}^A$, respectively, and their interrelation is mentioned in the text.}
		\label{fig:rt_hall_all}
	\end{figure*}

	As shown in Figs. {{\ref{fig:mt_mh_all}}}(a, b), the magnetization of the sample, in the whole temperature range, remains almost the same when a magnetic field ($\mu_0H$) is applied along the \textit{x}, and  \textit{y} axes (within the \textit{a-b} plane). In contrast to this, the magnetization of the sample along the \textit{z} axis is much smaller (Fig. {{\ref{fig:mt_mh_all}}}(c)). A clear signature of the two magnetic transitions is evident near 120 K and 242 K along all the axes. During all the field cooled (FC) measurements under 100 Oe of the applied magnetic field, large thermal hysteresis ($\sim$ 15 K) was observed near $T_{\text{N2}}$ while cooling (FCC) and warming (FCW) of the sample, as shown in Figs. \ref{fig:mt_mh_all}(a-c). Thermal hysteresis of nearly 10 K is also observed, near $T_{\text{N1}}$, in the FCC and FCW data as well. The thermal hysteresis remained the same when the entire measurement was performed at the rate of 2 K/min and 5 K/min, which suggests its intrinsic origin.

	In the case of \textit{M-T} measurements along the \textit{x} and \textit{y} axes, shown in Figs. {\ref{fig:mt_mh_all}}(a, b), small magnetization ($\sim$ 0.028(3) $\mu_{\text{B}}$/f.u.) is observed in the M1 regime at 130 K. However, the magnetization along \textit{z} axis remains much smaller, in the entire M1 regime, compared to the magnetization along the \textit{x} axis.
	The field dependent magnetization (\textit{M}(\textit{H})) at 130 K, along the \textit{x} and \textit{y} axes  (Figs. \ref{fig:mt_mh_all}(d, e)) shows a small residual magnetization near $\mu_0H=0$, and AFM (linear) behavior at higher fields. In the case of \textit{M}(\textit{H}) along the \textit{z} axis (at 130 K), the sample shows AFM behavior near zero fields and higher fields as well, as shown in Fig. {\ref{fig:mt_mh_all}}(f). The observation of tiny \textit{M}(\textit{H}) hysteresis at 130 K in the inset of Fig. {\ref{fig:mt_mh_all}}(f) is possible because of the imperfect alignment of the \textit{z} crystallographic axis of the crystal along the applied magnetic field. These observations suggest an in-plane canting of magnetic moments, leading to a small FM moment ($\sim$ 0.028(3) $\mu_{\text{B}}$/f.u. at 130 K) within the \textit{a-b} plane. Similar magnetization with a slightly lower residual moment ($\sim 0.018$ $\mu_{\text{B}}$/f.u.) has been observed in the case of Mn$_3$Ge as well {\cite{nayak2016large}}). Since the magnetization behavior of (Mn$_{0.78}$Fe$_{0.22}$)$_3$Ge in the M1 regime is similar to the magnetization behavior Mn$_3$Ge, it is possible that 22\% Fe doped Mn$_3$Ge compound (in M1 regime) also possesses an in-plane AFM structure similar to the Mn$_3$Ge. 
	
	Below $T_{\text{N2}}$ (M2 regime), small magnetization along all the axes has been observed, as shown in Figs. {\ref{fig:mt_mh_all}}(a-c). \textit{M}(\textit{H}) data of the sample, at 4 K and 85 K, show almost linear behavior (below 1 T) without hysteresis, as shown in Figs. {\ref{fig:mt_mh_all}}(d-f). These observations suggest the sample possesses  an AFM ordering at zero field in the M2 regime.  According to Ref. {\cite{hori1992antiferromagnetic}}, 22\% Fe doped Mn$_3$Ge compound possesses an AFM structure down to $\sim$70 K, followed by an onset of FM structure at a lower temperature. In our case, we observed a small splitting between the ZFC and FC curve, corresponding to all the axes, below 25 K  down to 2 K (Fig. \ref{fig:mt_mh_all}(a-c)). \textit{M}(\textit{H}) along the \textit{x} and \textit{y} axes, at 4 K, shows almost linear (antiferromagnetic) behavior below 1 T (insets of Figs. {\ref{fig:mt_mh_all}}(d-f)). However, it starts to saturate beyond 2 T. The nature of the \textit{M}(\textit{H}) curve is different at 4 K and 85 K, especially at the high magnetic fields ($\mu_0H>1$ T). These observations suggest the onset of field-induced small FM behavior in the sample below 25 K. However, at low field, the overall magnetization data suggest the dominance of AFM arrangements of the Mn+Fe moments in the entire M2 regime. 
	
	The observed magnetization in M1 and M2 regimes may arise from various possible magnetic structures. Therefore, we have performed neutron diffraction measurements of the sample (discussed later) to find magnetic models that describe the zero field magnetic structure of the sample in both  temperature regimes. 
	
	\section{Electrical transport results}
	The resistivity measurements of the sample, along all three high symmetry axes ($x, y, z$), display similar behavior in the temperature range of 2 K - 300 K, as shown in Fig. \ref{fig:rt_hall_all}(a). 	The nature of resistivity remains nearly the same along all three axes. The magnetic transitions are also visible near 120 K and 242 K in the resistivity measurements. Similar to magnetization, drastic changes in resistivity can also be observed near 120 K. Along with that, clear thermal hysteresis of $\sim$ 15 K is observed in the cooling-warming resistivity data along all three axes. The thermal hysteresis along the \textit{x} and \textit{z} axes, near 120 K, is shown in the inset of Fig. \ref{fig:rt_hall_all}(a). The sample shows an increase in resistivity with the decrease in temperature in the M1 regime. However, in the M2 regime, the resistivity shows metallic behavior along all three axes. This shows a strong coupling between the electronic properties and magnetism of the sample. 
	
	Hall resistivity measurements of the single-crystal (Mn$_{0.78}$Fe$_{0.22}$)$_3$Ge were performed with the magnetic field applied in and out of the \textit{x-y} plane (or \textit{a-b} plane). All the Hall resistivity data measured below $T_{\text{N1}}$ were antisymmetrized to nullify the magnetoresistance contribution due to the misalignment of the contacts. In the case of $\mu_0H$$\parallel$\textit{z}, no anomalous Hall resistivity was observed when the magnetic field approached zero (Fig. {\ref{fig:rt_hall_all}}(b)). However, in the case of current (\textit{I}) applied along the \textit{z} axis, and $\mu_0H$$\parallel$\textit{x}, a significant magnitude of anomalous Hall resistivity (AHR) was observed at 130 K, and 200 K (Fig. {\ref{fig:rt_hall_all}}(c)). The magnitude of AHR (denoted as $\rho_{zy}^{A}$) at a given temperature was determined by the linear fitting of the high field Hall resistivity data, as shown in Fig. {\ref{fig:rt_hall_all}}(c). Several Hall resistivity measurements were performed (for \textit{B}$\parallel$\textit{x})  in the temperature range of 4 - 300 K, and AHR at each temperature was determined. The temperature dependence of AHR ($\rho_{zy}^{A}$(\textit{T})) is shown in Fig. {\ref{fig:rt_hall_all}}(d), where it can be observed that the non-zero AHR is present only within the M1 regime.

	Usually, the Hall resistivity ($\rho_H$) can be defined as $\rho_H=R_0\mu_0H+R_MM$. Where $R_0$ and $R_M$ denote the ordinary Hall coefficient and Hall coefficient due to the magnetization of the sample, respectively. The paramagnetic nature of our sample at 300 K leads to the linear increase of the magnetization with an increase in the magnetic field $(M=\chi \mu_0H)$. Since the ordinary Hall resistivity also varies linearly with the magnetic field, $R_0$ and $R_M$ cannot be determined separately at 300 K. Therefore, the value of $R_0+R_M\chi$ at 300 K was found to be $0.043(2)$  ($\mu\Omega$ cm)/T. Using the high field Hall resistivity slope at 130 K, $R_0+R_M\chi$ was found to be $0.086(2)$  ($\mu\Omega$ cm)/T. At this temperature, for a magnetic field of 0.1 T, $(R_0+R_M\chi)\times(\mu_0H)$ gives  $0.0086(2)$  $\mu\Omega$ cm, which is small compared to the observed anomalous Hall resistivity ($0.74(1)$  $\mu\Omega$ cm). Moreover, similar to Mn$_3$Ge {\cite{kiyohara2016giant}},	the  $\rho_{zy}^{\text{Hall}}$  does not vary monotonically with the magnetization of the sample when magnetization is small (at low fields), as shown in the inset of Fig. {\ref{fig:rt_hall_all}}(c).		
		Therefore, the jump in AHR ($\rho_{zy}^{\text{Hall}}$), near zero fields, indicate the presence of an intrinsic anomalous Hall effect that does not arise due to residual net magnetization of the sample in the M1 regime. The most probable origin of this intrinsic Hall effect is the presence of non-vanishing Berry curvature, similar to the Mn$_3$Ge {\cite{kiyohara2016giant}}.
	
	The presence of finite Berry curvature gives rise to the intrinsic anomalous Hall conductivity (AHC). The AHC ($\sigma_{ij}^{A}$) corresponding to the observed anomalous Hall resistivity (AHR) can be calculated using the following relation: $\sigma_{ij}^{A}\approx-\rho_{ij}^{A}/(\rho_{ii}\rho_{jj})$. Where
	$\rho_{ii}$ and $\rho_{jj}$ are the longitudinal resistivities at the given temperature and magnetic field. $\rho_{ij}^{A}$ denote the AHR with current applied along the \textit{i} axis, voltage is measured along the \textit{j} axis, and the magnetic field is applied perpendicular to both the \textit{i} and \textit{j} axes. The temperature dependence of AHC (for  $\mu_0H$$\parallel$$x$), which was determined using the AHR at the corresponding temperature,  is shown in Fig. {\ref{fig:rt_hall_all}}(d). The magnitude of AHR, for $\mu_0H$$\parallel$\textit{x}, is nearly eight times smaller than the parent sample. Also, the longitudinal resistivity of the doped sample is up to six times larger than the parent sample {\cite{kiyohara2016giant}}. This leads to a nearly 50 times smaller magnitude of the AHC in our sample than Mn$_3$Ge {\cite{kiyohara2016giant}}. The interpretation of this observation will be discussed in the last section.
	
	Neutron diffraction analysis (to be discussed in the next section) shows that the magnetic structure of the sample in the M1 regime is noncollinear AFM, similar to Mn$_3$Ge {\cite{soh2020ground}}. We also observed that the AHC, similar to Mn$_3$Ge, is present within the M1 regime, which implies the presence of finite Berry curvature in this compound within the M1 regime. These observations underpin the possibility of the presence of Weyl points in (Mn$_{0.78}$Fe$_{0.22}$)$_3$Ge between $T_{\text{N2}}$ and $T_{\text{N1}}$.
	
	As we will conclude in the next section, the magnetic structure of the sample changes to collinear AFM in the M2 regime. The AHC vanishes in this region, which suggests that the Berry curvature also vanishes below  $T_{\text{N2}}$. The vanishing Berry curvature implies the absence of Weyl points in the M2 regime. The observation of vanishing AHC due to the change in the magnetic structure is reported in the case of Mn$_3$Sn as well {\cite{song2020complicated}}.
	Non-vanishing Berry curvature may be observed in collinear AFM systems with non-magnetic atoms residing on the non-centrosymmetric sites, creating asymmetric magnetization isosurfaces \mbox{\cite{vsmejkal2020crystal, vsmejkal2022anomalous, feng2020observation}}. However, in our case, the Ge (non-magnetic) atoms are located in a centrosymmetric  Wyckoff position. Therefore, Berry curvature is expected to vanish in the M2 regime, where the sample adopts a collinear AFM structure. The Weyl points (if exist) in a centrosymmetric system are separated in phase space, giving rise to the finite Berry curvature. Therefore, the vanishing Berry curvature in the M2 regime suggests the absence of Weyl points in the M2 regime.
	
	The underlying symmetry of the magnetic structure is the key to the existence of Weyl points and the resulting non-vanishing Berry curvature. Therefore, we have performed neutron diffraction of the single-crystal (Mn$_{0.78}$Fe$_{0.22}$)$_3$Ge to determine the magnetic structure of the sample in both temperature regimes.

	\begin{figure}[h] 
		\includegraphics[width=8cm]{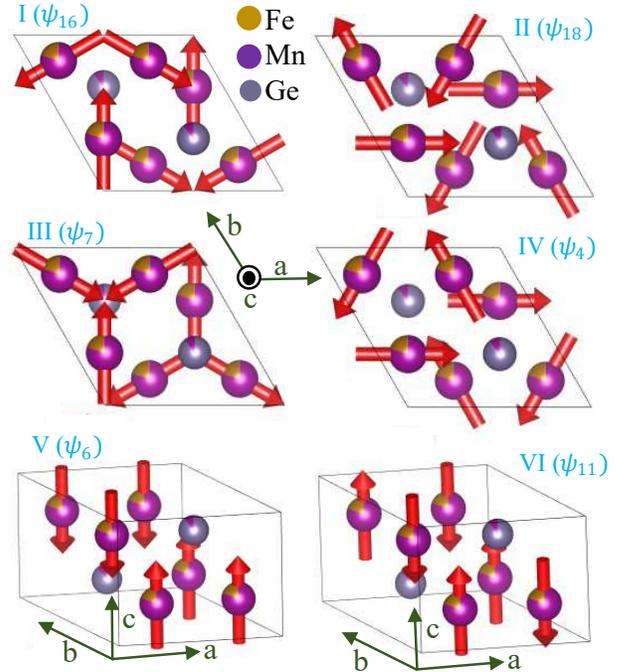}
		\caption{ (I - VI) Different magnetic models obtained from the representation analysis. Only models consistent with the magnetization of the sample are shown. $\psi_i$ denotes the basis vector. Here, the \textit{a} and \textit{b} axes lie along the [$2\bar{1}\bar{1}0$] and [$\bar{1}2\bar{1}0$] directions, respectively. In the case of the model (I, III), Mn+Fe moments point towards the  {[}$01\bar10${]} direction. However, in the case of models (II, IV), Mn+Fe moments point towards the  {[}$2\bar1\bar10${]} direction. Model (V, VI): AFM structures with moments lying along the \textit{c} axis ([0001]).}
		\label{fig:mag.structures}
	\end{figure}

	\section{Neutron diffraction}
	Neutron diffraction analyses have shown that Mn$_3$Ge possesses a coplanar noncollinear magnetic structure below 365 K \mbox{\cite{tomiyoshi1982magnetic, tomiyoshi1983triangular,nagamiya1982triangular, kren1970neutron}}. However, two symmetrically distinct candidate magnetic structures (Fig. {\ref{fig:mag.structures}}(I, II)) could not be differentiated using the unpolarized neutron diffraction technique. Finally, Ref. {\cite{soh2020ground}} used spherical neutron polarimetry to differentiate between these two structures. Since we also expect a magnetic structure similar to Mn$_3$Ge in the M1 regime of (Mn$_{0.78}$Fe$_{0.22}$)$_3$Ge, we performed neutron diffraction measurements of the sample using both the unpolarized and polarized neutrons. After combining the results of both the neutron diffraction experiments, the magnetic structure of (Mn$_{0.78}$Fe$_{0.22}$)$_3$Ge was determined in the M1 regime. The details of both experiments are described below.
	
	Since the magnetic form factors for the Mn and Fe are close to each other, it is difficult to separate the individual magnetic moments, which are located at the same Wyckoff position ($6h$).  Therefore, the entire analysis was performed with the assumption that both the Mn and Fe atoms at the $6h$ Wyckoff position are magnetically ordered. For convenience, we will refer to the combined moment of Mn and Fe as the Mn+Fe moment.
	
	Furthermore, if we assume that the Mn atoms occupying the Ge sites (referred to as Mn$_2$) are magnetically ordered, the refined moment value of  Mn$_2$ comes out to be small ($<$ 0.4(2) $ \mu_{\text{B}}$). Also, fitting parameters remain almost the same, irrespective of whether Mn$_2$ is considered to be magnetic or non-magnetic. Therefore, the analysis shown below is based on the assumption that the Mn atoms at Ge sites are non-magnetic.

	\subsection{Unpolarized neutron diffraction} \label{D23}
	Same as the parent sample, the propagation vector, $\mathbf{k}=0$ has also been reported for low Fe doped Mn$_3$Ge samples \cite{hori1992antiferromagnetic}. The single-crystal (unpolarized) neutron diffraction experiment was performed at different temperatures to determine the magnetic structures in the M1 and M2 regimes. The unpolarized neutron diffraction experiments were performed at 300 K, 130 K, 85 K, and 4 K, all of which were expected to show different behavior. The measurements at all the temperatures were performed for the scattering vector, $Q$ (= sin$\theta$/$\lambda$), ranging from 0.1 to 0.7 {\AA}$^{-1}$. The neutron diffraction data at all temperatures were analyzed using the JANA2006 software \cite{JANA2006}.
	
	\begin{table}[h]
		\caption{The refinement results of the D23 data analysis corresponding to the best fitted models at different temperatures. The refined lattice parameters and atomic positions at 300 K are also mentioned below.}
		
			\begin{tabular}{ccccc}
				\hline
				\hline
				Temperature & \multirow{1}{*}{4 K} & 85 K & 130 K & 300  K\tabularnewline[\doublerulesep]
				\hline 
				\noalign{\vskip\doublerulesep}
				Magnetic model & \multirow{1}{*}{Fig. \ref{fig:mag.structures}(V)} & Fig. \ref{fig:mag.structures}(V) & Fig. \ref{fig:mag.structures}(I; II) & - \tabularnewline[\doublerulesep]
				\hline 
				Moment ($\mu_{\text{B}}$) & 1.24(8) & 1.18(12) & 1.24(9); 1.25(8) & 0 \tabularnewline
				$\chi^{2}$  & 1.5 & 2.9 & 2.4 & 2.2 \tabularnewline
				\multicolumn{1}{c}{R$_{w}$}  & 4.3 & 3.9 & 3.1 & 2.8 \tabularnewline[\doublerulesep]
				Domain ratio & - & - & 25:39:41; 42:31:29 & - \tabularnewline[\doublerulesep]
				No. of independent\\ reflections & 25 & 19 & 25 & 22 
				\tabularnewline[\doublerulesep]
				\hline	
			\end{tabular}
			\\
			\begin{tabular}{c}
				
				At $T = 300$ K,	$a=b=5.300(1)$ \AA, $c=4.271(1)$ \AA
				\tabularnewline[\doublerulesep]
				\hline

			\end{tabular}		
			\\
			\begin{tabular}{cccc}
				\hline
				Atom (site) & \textit{x} & \textit{y} & \textit{z}
				\tabularnewline[\doublerulesep]
				\hline 
				\noalign{\vskip\doublerulesep}
				Mn/Fe ($6h$) & 1/3 & 2/3 & 0.25 
				\tabularnewline[\doublerulesep]
				Ge ($2c$) & 0.8335(2) & 0.6670(4) & 0.25 
				\tabularnewline[\doublerulesep]
				\hline
				\hline
				
			\end{tabular}
			\label{models_best}		
	\end{table}

\begin{figure} [h]
		\includegraphics[width=8cm]{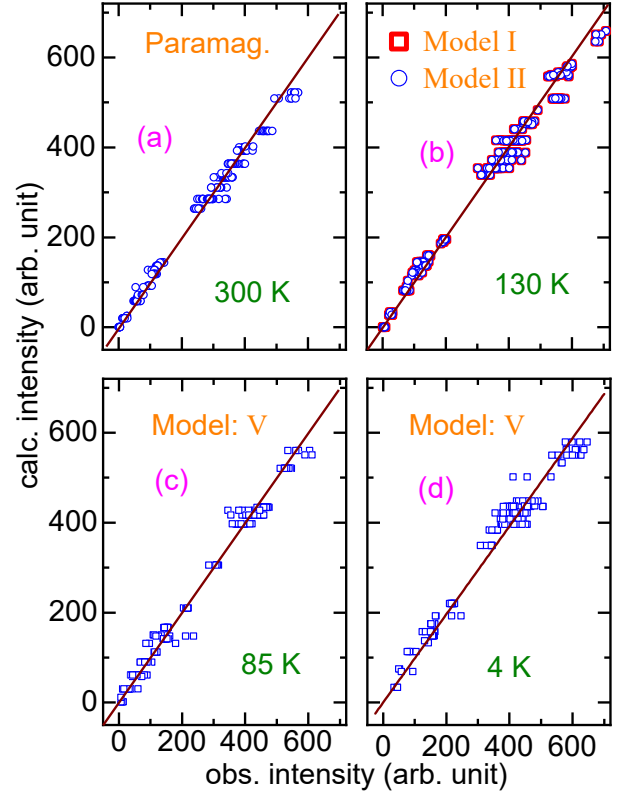}
		\caption{ (a-d) Observed (obs.) integrated intensities of the Bragg reflections are plotted against the calculated (calc.) values at 300 K, 130 K, 85 K, and 4 K, respectively. The Lorentz factor has already been corrected in the observed intensities. In the case of 4 K, 85 K, and 130 K, the data correspond to total intensities (magnetic + nuclear). At 300 K, the sample is paramagnetic (Paramag.), so the data correspond to nuclear intensity only. The magnetic model corresponding to the fitted data is mentioned in each plot, and the magnetic models are shown in  Fig. \ref{fig:mag.structures}.}
		\label{fig:Unpol_Obs-calc}
	\end{figure}

	As mentioned in section {\ref{exp_detail}}, initially, the neutron diffraction experiment was performed at 300 K. The integrated intensities corresponding to 410 Bragg reflections were collected, and the data were analyzed using the $P6_3/mmc$ space group (No. 194) symmetry. We observed a good agreement between the calculated and observed intensities at 300 K, as shown in Fig. {\ref{fig:Unpol_Obs-calc}}(a). The obtained goodness-of-fit statistics ($\chi^2$ and $\text{R}_w$)  are mentioned in Table {\ref{models_best}}. The Fe doping fraction and occupancy of different sites were also refined at this temperature.  The obtained chemical formula can be written as (Mn$_{0.79(2)}$Fe$_{0.21(2)}$)$_{3}$Mn$_{0.12(5)}$Ge$_{0.88(5)}$. This suggests that $\sim$ 12\% Ge sites (2\textit{c} Wyckoff position) are occupied by Mn atoms, which is slightly higher than expected from the chemical analysis ($\sim 6$ \%). It also implies the presence of a small amount of ($\sim 1.5 - 2\%$)  unreacted Mn atoms, which are sitting randomly in the sample. Moreover, we observed that $\sim$ 21\% of the Mn sites ($6h$ Wyckoff position) are occupied by Fe atoms, which are near the expected values.  During the analysis, $\chi^2$ slightly increases when Fe atoms are allowed to occupy the Ge sites ($2c$ Wyckoff position), which suggests that the Fe atoms might not be occupying the Ge sites.

	\begin{table}[h]
		\caption{The goodness-of-fit statistics ($\chi^2$, R$_w$) obtained from the D23 experiment data fitted with the various possible models at 130 K. The mentioned (magnetic) models are shown in Fig. {\ref{fig:mag.structures}}}
		\begin{tabular}{ccccccc}
			\hline
			\hline
			(130 K) Model & \multirow{1}{*}{I } & II & III & IV &  V & VI  \tabularnewline[\doublerulesep]
			\hline 
			\noalign{\vskip\doublerulesep}
			$\chi^{2}$  & 2.4 & 2.4 & 5.5 & 4.6 & 3.9 & 3.2
			\tabularnewline[\doublerulesep]
			\multicolumn{1}{c}{R$_{w}$}  & 3.1 & 3.1 & 6.5 & 5.4 & 4.1 & 3.5 \tabularnewline[\doublerulesep]
			\hline
			\hline
		\end{tabular}
		\label{tab:D23_130K}
	\end{table}

	Based on the neutron powder diffraction of the (Mn$_{0.82}$Fe$_{0.18}$)$_{3.25}$Ge compound, the propagation vector $\mathbf{k}=0$ has been reported in the M1 and M2 magnetic regimes {\cite{hori1992antiferromagnetic}}, which implies that the magnetic and nuclear Bragg reflections coincide. In our case, the neutron diffraction data at 130 K, 85 K, and 4 K show that the integrated intensities at nuclear Bragg positions are significantly higher than the integrated intensities at 300 K, which reconfirms the reported propagation vector $(\mathbf{k}=0)$. The refined site occupancy and atomic positions obtained by the 300 K data analysis were kept fixed during the low temperature data analysis. However, the Debye-Waller factor for each atom was refined at low temperatures.
	
	We used representation analysis to determine symmetry allowed possible magnetic structures of the sample {\cite{wills2001magnetic}}.  SARA\textit{h}-Representational Analysis {\cite{saraha}} shows that for $\mathbf{k}=0$ and the space group $P6_3/mmc$, the magnetic representation for the Mn site can be decomposed into ten irreducible representations (IRs). The basis vectors associated with these IRs are denoted as $\psi_i$ (\textit{i} = 1 to 18). These IRs correspond to various possible magnetic models, including ferromagnetic, ferrimagnetic, and antiferromagnetic. As mentioned in section {\ref{magnetization}}, the magnetization of the sample at 130 K suggests an AFM behavior of the sample. Therefore, the basis vectors which represent ferromagnetic/ferrimagnetic models can be disregarded from the possible magnetic structures at 130 K. These ferromagnetic/ferrimagnetic models were disregarded based on our neutron data analysis as well. The AFM models compatible with the magnetization of the sample are shown in Fig. {\ref{fig:mag.structures}} (I - VI). These magnetic models were fitted with the D23 data at 130 K, keeping the occupancy of Mn and Ge sites fixed as obtained by the neutron data analysis at 300 K. The magnetic moment, domain ratio, and (isotropic) Debye-Waller factors were refined for all the AFM models shown in Fig. {\ref{fig:mag.structures}}. The goodness-of-fit statistics corresponding to various models are shown in Table {\ref{tab:D23_130K}}, where it can be observed that models I and II fit best at 130 K. The magnetic intensity corresponding to models I and II were found to be the same, which is expected \cite{brown1990determination}. The calculated vs. observed intensities corresponding to these two models are shown in Fig. \ref{fig:Unpol_Obs-calc}(b), where it is evident that they cannot be distinguished. The refinement results corresponding to the two best possible models (I, II) at 130 K are shown in Table \ref{models_best}. The magnetic structure of Mn$_3$Ge (model I) has been determined by Ref. {\cite{soh2020ground}} using the spherical neutron polarimetry (SNP) technique. Therefore, we have also employed the SNP technique to distinguish between models I and II and found the ground state magnetic structure of (Mn$_{0.78}$Fe$_{0.22}$)$_3$Ge at 130 K. The details of the SNP experiment and results will be discussed in the next section (\ref{cryopad}).
	
	\begin{table}
		\caption{The goodness-of-fit statistics ($\chi^2$; R$_w$) obtained from the D23 data (at 4 K, 85 K) fitted with all the models shown in Fig. {\ref{fig:mag.structures}}.}		
		\begin{center}
			\begin{tabular}{c c c  c } 
				\hline\hline
				Model &  $T=$ 4 K ($\chi^2$; R$_w$) & & $T=$ 85 K ($\chi^2$; R$_w$) \\ [0.5ex] 
				\hline
				Fig. \ref{fig:mag.structures}(I) & 3.2; 11.7 & & 17.4; 21.3 \\ 
				Fig. \ref{fig:mag.structures}(II)  & 3.2; 11.9 & & 17.2; 21.2 \\
				Fig. \ref{fig:mag.structures}(III)  & 5.4; 12.7 & & 26.1; 29.5 \\
				Fig. \ref{fig:mag.structures}(IV)  & 4.9; 11.7 & & 21.2; 25.1 \\  
				Fig. \ref{fig:mag.structures}(V) & 1.5; 4.3 & & 2.9; 3.9 \\
				Fig. \ref{fig:mag.structures}(VI) & 3.3; 10.5 & & 14.5; 17.1 \\
				\hline
				\hline
			\end{tabular}
		\end{center}
		\label{tab:D23_4K_85K}
	\end{table}

	As mentioned earlier, the propagation vector at 4 K and 85 K remain the same, i.e. $\mathbf{k}=0$. Therefore, the same 18 basis vectors associated with ten IRs (as mentioned above) were obtained at these temperatures as well. In this temperature regime (M2), the magnetization shows AFM behavior, which suggests that all the IRs corresponding to ferromagnetic/ferrimagnetic structures can be disregarded. However, the AFM models among  I - VI (Fig. {\ref{fig:mag.structures}}) may be the magnetic structure of the sample at 4 K and 85 K. Therefore, we have refined neutron diffraction data at both temperatures, using all the six magnetic models shown in Fig. {\ref{fig:mag.structures}}. Similar to the data analysis at 130 K, site occupancy and atomic position of the atoms were kept the same as obtained at 300 K. Only the Debye-Waller factors and magnetic moment were refined at 4 K and 85 K. The goodness-of-fit statistics ($\chi^2$; R$_w$) corresponding to models (I-VI) are mentioned in the Table \ref{tab:D23_4K_85K}. It is evident from Table \ref{tab:D23_4K_85K} that model V fits significantly better than any other model at 4 K and 85 K. Therefore, it can be concluded that the magnetic structure of the sample at both temperatures is the same as presented in Fig. \ref{fig:mag.structures} (V), which corresponds to the Shubnikov space group $P6_3^{'}/mm{'}c$ ($\psi_{6}$). Our observation reconfirms the conclusion made by Ref. \cite{hori1992antiferromagnetic}, which also suggests model V to be the magnetic structure of the sample in the M2 regime. The calculated vs. observed intensities (at $T=$ 85 K and 4 K) corresponding to the model V are shown in Figs. \ref{fig:Unpol_Obs-calc}(c, d), and respective relevant fitted parameters are mentioned in Table \ref{models_best}. The obtained Mn+Fe moment value ($\sim$1.24(8) $\mu_{\text{B}}$) is slightly lower than that obtained by Ref. {\cite{hori1992antiferromagnetic}} (1.4 $\mu_{\text{B}}$), which is possible because the sample used by Ref. {\cite{hori1992antiferromagnetic}} has slightly lower Fe doping than our sample.
	
	\begin{figure}[h] 
		\includegraphics[width=7.5cm]{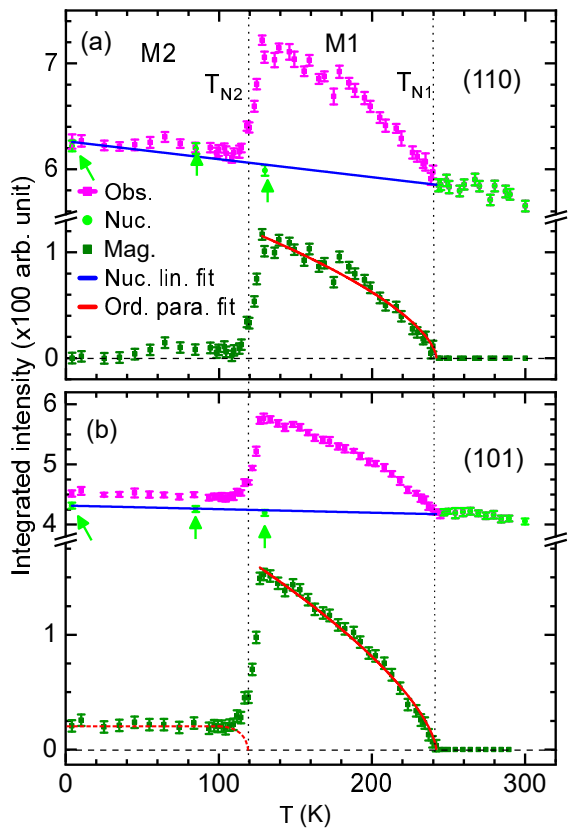}
		\caption{(a, b) shows the temperature dependence of the (110) and (101) reflections, respectively, determined by the single-crystal neutron diffraction using the D23 instrument. Obs. denotes the observed intensity. Nuc. and Mag. denote the nuclear and magnetic intensities, respectively. The light green arrows at 4 K, 85 K, and 130 K denote the nuclear contribution  determined by the neutron diffraction analysis at these temperatures. Nuc. lin. fit denotes the linear fitting of the nuclear contribution at 4 K, 85 K, 130 K, and 242 K. Ord. para fit denotes the fitting of the curve in the M1 regime using the order parameter relation mentioned in the text. Since the sample is paramagnetic beyond 242 K, nuclear intensity above this temperature is the same as the observed intensity. The dashed red line in (b), below 100 K, is a guide to the eye for the order parameter in the M2 regime.}
		\label{fig:neutron-T}
	\end{figure}
	
	The temperature dependence of the integrated intensities of (110) and (101) reflections were measured (while heating the sample), and is shown in Fig. \ref{fig:neutron-T}. The observed intensities of (110) and (101) reflections remain almost the same with the increase in temperature up to $T_{\text{N2}}$. The intensity of both the reflections suddenly increases beyond $T_{\text{N2}}$ and reaches a maximum near 125 K. Beyond this temperature, in the M1 regime, the intensity decreases up to $T_{\text{N1}}$ and remains nearly constant beyond that. 
	
	Using the neutron diffraction analysis at 4 K, 85 K, and 130 K,  nuclear and magnetic contributions were separated at these temperatures. 
  Since the sample is paramagnetic beyond 242 K, the measured integrated intensities at 242 K and above can be regarded as purely nuclear. As shown in Fig. {\ref{fig:neutron-T}}, the nuclear intensity slightly decreases with an increase in the temperature because of the Debye-Waller factor.  The nuclear intensity contribution between 4 - 242 K was determined by linearly fitting the nuclear intensities at 4 K, 85 K, 130 K, and 242 K. Further, the magnetic intensities between 4 - 242 K were extracted by subtracting the nuclear intensities from the total intensity. It has been observed that the magnetic contribution corresponding to (110) reflection remains negligible in the M2 regime (Fig. \ref{fig:neutron-T}(a)).  In contrast to this, a significant magnitude of magnetic intensity is present in the (101) reflection within the M2 regime (Fig. {\ref{fig:neutron-T}}(b)). It remains almost constant with temperature between 4 K to $\sim100 $ K. This is consistent with our previous observation that the magnetic moment of Mn+Fe remains nearly the same at 4 K and 85 K (Table \ref{models_best}). These observations suggest that the order parameter in the M2 regime remains constant with temperature and vanishes near 120 K, as illustrated by the dashed red line in Fig. \ref{fig:neutron-T}(b).
	
	As shown in  Fig. {\ref{fig:neutron-T}}, significant magnetic intensity is present for the (101) and (110) reflections within the M1 regime. In this temperature range, the magnetic intensity decreases with an increase in temperature and vanishes near $T_{\text{N1}}$. Magnetic intensity (denoted as $I_{m}(T)$), in the M1 regime, follows the order parameter relation: $I_{m}(T)=I_{m}^o(1-T/T_{\text{N1}})^{2\beta}$. Here, $I_{m}^o$ is the magnetic intensity at $T=0$ and $\beta$ is an exponent, which was found to be 0.31(2) and 0.33(1) for the (110) and (101) reflections, respectively. The observed exponent value in our case is somewhat larger compared to Mn$_3$Ge ($\beta=0.21$) \cite{chen2020antichiral}. 
	
	\subsection{Polarized neutron diffraction} \label{cryopad}
	The unpolarized neutron diffraction could not differentiate between the magnetic models I and II shown in Fig. \ref{fig:mag.structures}. 
		Therefore, similar to Ref. {\cite{soh2020ground}}, we also performed polarized neutron diffraction of the sample using the CRYOPAD setup. We utilized the spherical neutron polarimetry (SNP) technique to distinguish between these models, and determined the ground state magnetic structure of the sample in the M1 regime. In addition, we performed SNP measurement at 4 K as well to reconfirm the magnetic structure determined using the D23 instrument.

	\begin{figure}[h] 
		\includegraphics[width=8.5cm]{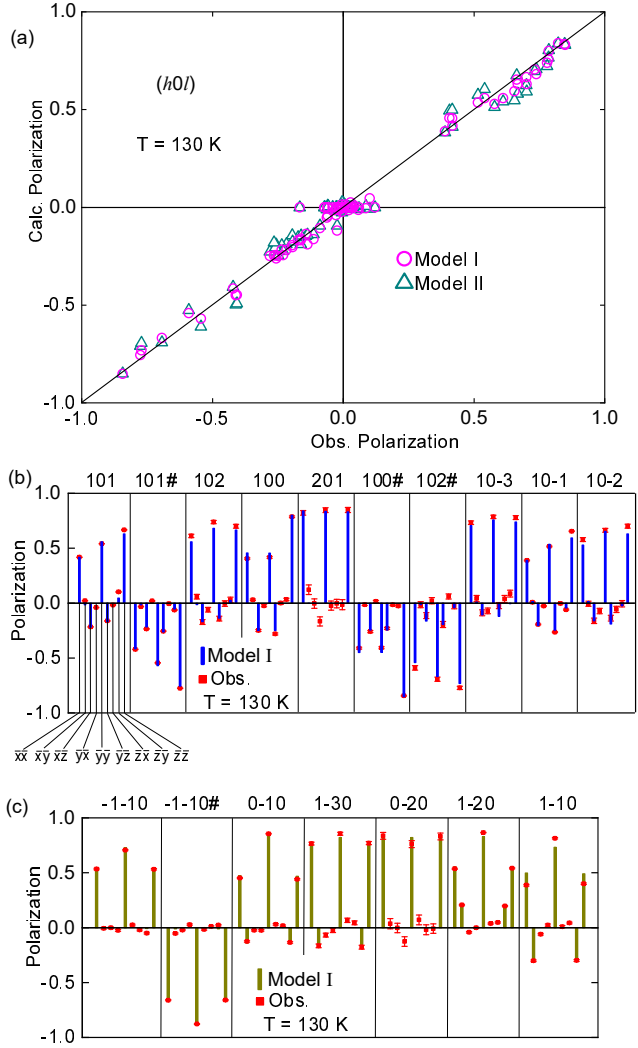}
		\caption{(a) Calculated (Calc.) vs. observed (obs.) polarizations corresponding to the ($h0l$) reflections. (b, c) Calculated and observed polarization matrix elements ($P_{ij}$ ($i,j=\bar{\mathsf{x}}, \bar{\mathsf{y}}, \bar{\mathsf{z}}$)) at 130 K, corresponding to the ($h0l$) and ($hk0$) reflections, respectively. The $\bar{\mathsf{x}}\bar{\mathsf{x}}, \bar{\mathsf{x}}\bar{\mathsf{y}}...\bar{\mathsf{z}}\bar{\mathsf{z}}$ mentioned in (b) denote $P_{\bar{\mathsf{x}}\bar{\mathsf{x}}}, P_{\bar{\mathsf{x}}\bar{\mathsf{y}}}$...$P_{\bar{\mathsf{z}}\bar{\mathsf{z}}}$, respectively. The reflections corresponding to each set of polarization are mentioned at the top. The reflections followed by \# symbols suggest that the measurement was performed with the reversed incident polarization.}
		\label{fig:polarization_bars_130K}
	\end{figure}
	
	\begin{figure}[h] 
		\includegraphics[width=8.5cm]{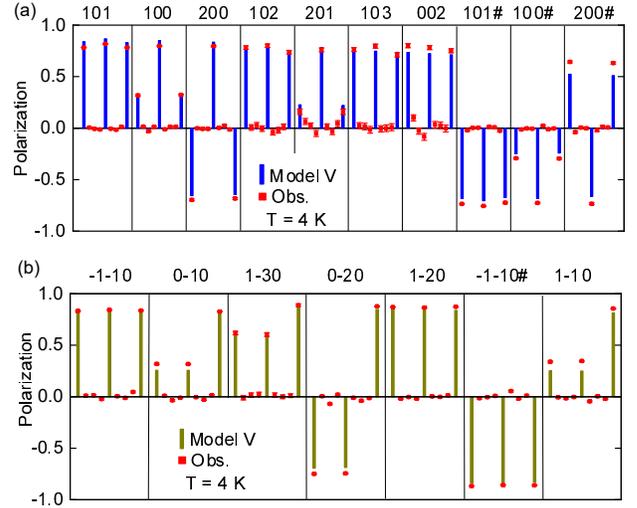}
		\caption{Calculated and measured $P_{ij}$ at 4 K, corresponding to the (a) ($h0l$), and (b) ($hk0$) reflections. Same as previous, the reflections followed by \#  were measured with the reversed incident polarization.}
		\label{fig:polarization_bars_4K}
	\end{figure}

	In the M1 and M2 regimes, the experiments were performed at 130 K and 4 K, respectively. The polarization matrix elements ($P_{ij}$) corresponding to the various ($h0l$) and ($hk0$) reflections were measured. The obtained data were fitted using the \textsc{Mag2Pol} software \mbox{\cite{qureshi2019mag2pol, qureshi2018magnetic}}. During the analysis, the polarization data were corrected for spin filter efficiency, and initial polarization (0.935) was also taken into account. The structural parameters, such as lattice parameters, atomic positions, lattice symmetry, and site occupancy, used for the CRYOPAD data analysis were obtained by the D23 data refinement at 4 K and 130 K. During the CRYOPAD data refinement, only the magnitude of magnetic moment and the fraction of the magnetic domains were refined, keeping all structural parameters as constants.
	
	The data obtained at 130 K were fitted with different models shown in Fig. {\ref{fig:mag.structures}}(I-VI), and their goodness-of-fit statistics, $\chi_r^2$, is compared in Table {\ref{cryopad_diff_models}}. The comparisons for the ($h0l$) reflections show that model I has a significantly smaller $\chi_r^2$ compared to any other model. The calculated vs. observed polarizations for ($h0l$) reflections,  corresponding to models I and II, are also shown in Fig. \ref{fig:polarization_bars_130K}(a). It can be observed that most of the calculated polarizations corresponding to model I fit better than those calculated corresponding to model II. In contrast to the ($h0l$) reflections, ($hk0$) reflections show a negligible difference in the goodness-of-fit statistics when models I and II  were compared (Table \ref{cryopad_diff_models}). These observations suggest that the magnetic model I represents the magnetic structure of the sample in the M1 regime.  The observed and calculated polarization matrix elements ($P_{ij}$) for the best-fitted model (model I) are shown in Fig. \ref{fig:polarization_bars_130K}(b, c), where it can be seen that the calculated polarization matrix elements match with the observed matrix elements for most of the reflections. 
	
	The refined parameters corresponding to the model I, fitted with 130 K data, are shown in Table {\ref{cryopad_best}}.	Usually, neutron polarimetry is not sensitive to the magnitude of the magnetic moment. However, it can be determined in our case because of the $\mathbf{k}=0$, which leads to strong nuclear and magnetic interference scattering \mbox{\cite{soh2020ground, chatterji2005neutron}}. The average magnetic moment (average from ($hk0$) and ($h0l$) reflections) determined by the fitting of the model I at 130 K is 1.4(1) $\mu_{\text{B}}$, which is the same (within the error bar) as obtained by the D23 analysis (1.24(9) $\mu_{\text{B}}$). This also strengthens the fact that the magnetic structure of the 22\% Fe doped Mn$_3$Ge compound remains the same as Mn$_3$Ge \cite{soh2020ground} (Model I) in the M1 regime.
	
	\begin{table}[h]
		\caption{The goodness-of-fit statistics corresponding to the CRYOPAD experiment data fitted with the different possible models at 130 K and 4 K.}
		
		\begin{tabular}{ccccccc}
			\hline
			\hline
			Model (\textit{T} = 130 K) & \multirow{1}{*}{I } & II & III &  IV & V & VI \tabularnewline[\doublerulesep]
			\hline 
			\noalign{\vskip\doublerulesep}
			$\chi^{2}_r$ ($h0l$) & 15 & 78 & 1934 & 943 & 1784 & 1729
			\tabularnewline[\doublerulesep]
			\multicolumn{1}{c}{$\chi^{2}_r$ ($hk0$)} & 19 & 20 & 398 & 489 & 314 & 348  
			\tabularnewline[\doublerulesep]
			\hline
		\end{tabular}
		\\
		
		\begin{tabular}{ccc}
			\hline
			Model (\textit{T} = 4 K) & \multirow{1}{*}{Fig. \ref{fig:mag.structures}(V) } &  Fig. \ref{fig:mag.structures}(VI)\tabularnewline[\doublerulesep]
			\hline 
			\noalign{\vskip\doublerulesep}
			$\chi^{2}_r$ ($h0l$)  & 22 & 913 
			\tabularnewline[\doublerulesep]
			\multicolumn{1}{c}{$\chi^{2}_r$ ($hk0$)}  & 47 & 6998 \tabularnewline[\doublerulesep]
			\hline
			\hline
			
		\end{tabular}
		\label{cryopad_diff_models}
	\end{table}
	
	Since the unpolarized neutron diffraction analysis did not show any difference in magnetic structure at 4 K and 85 K, we performed the polarimetry experiment at 4 K only. Similar to 130 K, all the nine matrix elements were measured at 4 K, corresponding to several ($hk0$) and ($h0l$) reflections. The observed polarization matrix elements were fitted with the two most probable AFM structures (model - V, VI), and respective  $\chi_r^2$ is mentioned in Table \ref{cryopad_diff_models}, where it is obvious that model V fits much better than model VI. This supports our previous conclusion (based on the D23 analysis) that model V is the magnetic structure of the sample at 4 K. The fitting parameters corresponding to model V are mentioned in Table \ref{cryopad_best} (4 K). The individual values of observed and calculated $P_{ij}$ corresponding to the best-fitted model at 4 K are shown in Fig. \ref{fig:polarization_bars_4K}. The average magnetic moment value determined at this temperature is 1.5(1) $\mu_{\text{B}}$, which is slightly higher than those obtained by the D23 analysis (1.24(8) $\mu_{\text{B}}$). Since the propagation vector, $\mathbf{k}=0$ for our compound, nuclear parameters cannot be accurately determined using the D23 analysis. Also, as mentioned previously, the moment value determined by the CRYOPAD analysis is sensitive to nuclear parameters (due to the nuclear-magnetic interference). Therefore, such a small difference in magnetic moment value determined by two different analyses is possible.
	
	\begin{table}[h]
		\caption{The refinement results of the CRYOPAD experiment for best-fitted models at 4 K and 130 K. Here, ($h0l$) and ($hk0$) correspond to the experimental setup in which the \textit{b} and \textit{c} axis, respectively, of the sample were kept in the vertical position.}
		
		\begin{ruledtabular} %
			\begin{tabular}{ccc}
				Temperature & \multirow{1}{*}{4 K [($h0l$), ($hk0$)]} & 130 K [($h0l$), ($hk0$)] \tabularnewline[\doublerulesep]
				\hline 
				\noalign{\vskip\doublerulesep}
				Best model & \multirow{1}{*}{Fig. \ref{fig:mag.structures}(V)} & Fig. \ref{fig:mag.structures}(I)  \tabularnewline[\doublerulesep]
				\hline 
				Moment ($\mu_{\text{B}}$) & 1.44(6), 1.59(9) & 1.42(6), 1.38(7) 
				\tabularnewline
				$\chi^{2}_r$  & 22, 47 & 15, 19 
				\tabularnewline[\doublerulesep]
				Domain ratio & - & 16:30:54, 60:25:15  \tabularnewline[\doublerulesep]
				No. of $P_{ij}$ elements  & 90, 78 & 90, 72  \tabularnewline[\doublerulesep]
		\end{tabular}\end{ruledtabular} \label{cryopad_best}
	\end{table}

	In conclusion, the magnetic structure of the 22\% doped Mn$_3$Ge compound in the M1 regime remains the same as the parent sample - Mn$_3$Ge. In the M2 regime, the sample possesses a collinear AFM structure with Mn+Fe moments lying along the \textit{c} axis.  
	
	\section{CONCLUSION}
	Magnetization and resistivity measurements of (Mn$_{0.78}$Fe$_{0.22}$)$_3$Ge show magnetic phase transitions near $T_{\text{N1}}$ and $T_{\text{N2}}$. The magnetization of the sample in the M1 regime shows similarity with the magnetization of Mn$_3$Ge. The magnetic structures of the sample were determined using single-crystal unpolarized neutron diffraction and spherical neutron polarization analysis techniques. Interestingly, the magnetic structure of (Mn$_{0.78}$Fe$_{0.22}$)$_3$Ge in the M1 regime is a triangular AFM in the \textit{a-b} plane with the Mn+Fe moments arranged the same way as the parent compound - Mn$_3$Ge. In the M2 regime, the sample possesses a collinear AFM structure with Mn+Fe moments parallel to the \textit{c} axis. 
	
	Non-zero AHC is expected and observed in the M1 regime. The magnetic structure of the sample also remains the same as Mn$_3$Ge within this temperature regime.  Therefore, Weyl points are likely to be present in (Mn$_{0.78}$Fe$_{0.22}$)$_3$Ge between $T_{\text{N2}}$ and $T_{\text{N1}}$. The observed AHC in our sample is nearly 50 times smaller than the parent compound - Mn$_3$Ge. The separation and location of the pair of Weyl points near the Fermi surface determine the magnitude of AHC \cite{wuttke2019berry}. This suggests that Fe doping in Mn$_3$Ge led the pair of Weyl points to move farther away from the Fermi surface, compared to their location in the parent phase - Mn$_3$Ge. AHC is present in the M1 regime. However, it is absent in the M2 regime. Therefore, it can be inferred that the characteristics of Weyl points are interlinked with the magnetic structure of the material.
	
	Finally, we can conclude that the Fe doping in Mn$_3$Ge can tune the characteristics of the Weyl points present in the sample. However, detailed transport measurements and analysis of several doped samples will be required to validate our conclusions regarding the characteristics of the Weyl points. The likelihood of the presence of Weyl points in the doped sample suggests the robust nature of the topology of the electronic bands in Mn$_3$Ge. These features of a magnetic Weyl semimetal are promising for its future application.
	
	\section{Acknowledgement}
	The unpolarized neutron diffraction experiment was performed at ILL, France using the D23 setup operated as a CRG instrument by CEA Grenoble. The proposal number for this experiment is CRG-2798. The polarized neutron diffraction experiment was performed using the CRYOPAD setup (proposal number 5-54-361 \cite{doi_cryopad}).

	\bibliographystyle{apsrev} 
	
		\bibliography{ref_22Fe_Mn3Ge_Neut}

\end{document}